\documentstyle[seceq]{ptptex}

\input epsf

\title{
M-Theory and the Light Cone
}

\author{
Joseph {\sc Polchinski}\footnote{joep@itp.ucsb.edu}
}

\inst{
Institute for Theoretical Physics,\\
University of California, Santa Barbara, CA 93106-4030
}

\abst{I discuss D0-brane quantum mechanics as a
nonperturbative formulation of string theory, in particular the
relation between the Banks--Fischler--Shenker--Susskind matrix
model, the Maldacena conjecture for
D0-branes, and type IIA/M-theory duality.  Some features of the
quantum mechanics of D0-branes are also discussed.
}

\begin{document}

\maketitle

\section{Introduction}

The main focus of this lecture is D0-brane quantum mechanics as a
nonperturbative formulation of string theory.  In particular I
would like to discuss the relation between the
Banks--Fischler--Shenker--Susskind matrix model,\cite{BFSS} the
Maldacena conjecture for D0-branes\cite{malda,IMSY}, and type
IIA/M-theory duality.\cite{IIAM}  This
has sometimes been a confusing subject, and it is one of which
everyone has their own understanding.\footnote
{Other discussions of which I am aware, some of which overlap mine,
are listed in ref.~\cite{others}.}  I hope that I will reduce rather
than add to the confusion.  I have certainly added to it in the
past; the present version reflects my improved understanding.

Figure~1 is a schematic picture of the lessons of string duality,
\begin{figure}
\begin{center}
\leavevmode
\epsfbox{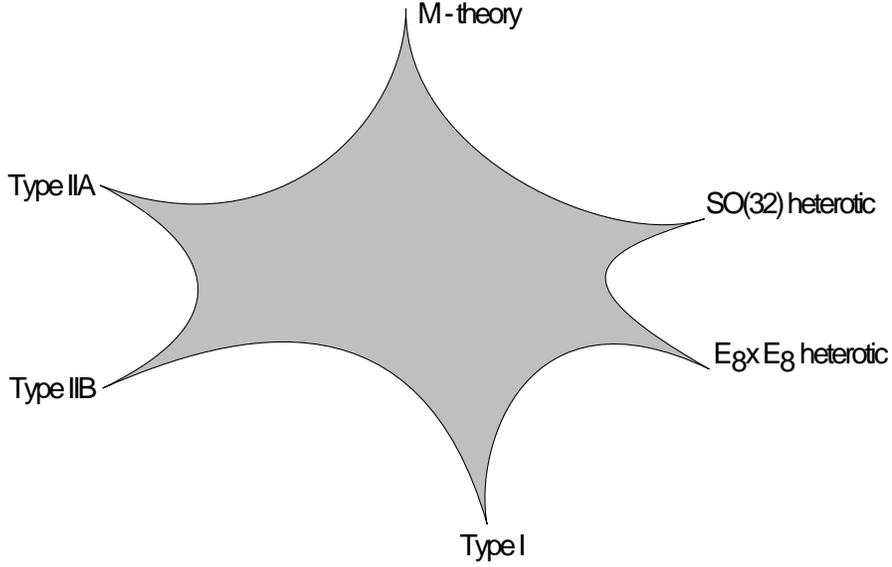}
\caption{Space of string vacua.  The cusps are limits in which a
weakly coupled string description is possible (except for the
M-theory limit).}
\end{center}
\end{figure}
one that I have drawn perhaps a hundred times.
It is intended to emphasize the limited range of conventional
string theory, which only provides an asymptotic expansion at the
five stringy cusps.  This is true even
when one includes D-branes --- only asymptotically in the string
coupling do these provide a quantitatively accurate description.  To
describe the complete physics at a finite distance from a cusp, and
to reach the M-theory point with 11-dimensional Lorentz invariance,
requires a leap, a conjecture, to a new idea with a greater range of
validity.  This new idea can only be guessed, not deduced from
information already in string theory.

This has always been the intended message of figure~1, but as I
will explain I now believe, rather surprising, that one can actually
deduce a great deal from what is already known. The main outline of
this talk is the description of four limits: the DKPS limit, the SSS
limit, the BGLHKS limit, and the IMSY limit. Actually, these are all
the same limit of type IIA string theory, but presented in different
ways and with different emphases.  I will then explain how these
different descriptions illuminate the relations among the various
ideas.  At the end I discuss possible future directions in D0-brane
quantum mechanics, and some properties of the D0-brane
bound state.
\medskip

\section{The Douglas--Kabat--Pouliot--Shenker (DKPS)
Limit}

Consider a state in the IIA string theory containing $N$
D0-branes plus any collection of open and closed
strings.\footnote{For reviews of D0-brane properties see
refs.~\cite{dbranes}.}  The energy of such a state is
\begin{eqnarray}
H &=& \frac{N}{g_{\rm s} l_{\rm s}} + O(l_{\rm s}^{-1}) + O(q)
\nonumber\\ && {\hspace{-15pt}
}+ \frac{g_{\rm s}l_{\rm s}}{2} {\rm Tr}(P^2) - \frac{1}{16\pi^2
g_{\rm s}  l_{\rm s}^5}
\sum_{m,n} {\rm Tr} ([X^m,X^n]^2) + \mbox{spin term\,+\,higher
derivs}.\qquad\quad \label{hamilt}
\end{eqnarray}
Here $g_{\rm s} $ is the dimensionless string coupling and $l_{\rm
s} = (\alpha')^{1/2}$ is the string length scale.  The first term
is the D0-brane rest energy.  The second is the mass of excited
open or closed strings.  The third is from massless closed strings,
with momenta of order~$q$.

The important terms are in the second line.  The first is a
nonrelativistic kinetic term, $P^2/2m$, except that it is a trace
over $N^2$ terms rather than a sum over $N$ terms.  This is a
reflection of the important non-Abelian geometry of
D-branes.\cite{witbound}  The coordinates arise from
the lightest open string state
\begin{equation}
\psi^m_{-1/2} | ij \rangle\ .
\end{equation}
This state has a 9-dimensional spatial index $m$ and so naturally
corresponds to a collective coordinate.
The ket $| ij \rangle$ denotes a string that begins on the
$i$th D0-brane and ends on the $j$th, so the coordinate
$X^m_{ij}$ is a {\it matrix} in the $N$-dimensional D0-brane index
space.  Then $P^m_{ij}$ is the conjugate momentum.  The potential
term corrects this overcounting.  It is nonnegative (the minus
sign canceling an $i^2$) and vanishes precisely when all the
$X^m$ commute and so can be diagonalized.  Further, there is a
$U(N)$ gauge invariance, so the state is independent of basis
and the low energy physics reduces to the quantum mechanics of
the $9N$ eigenvalues, as appropriate for $N$ particles in $9+1$
dimensions.

The potential term has the same form as the 4-gluon Yang--Mills
interaction Tr$(F^{mn} F_{mn})$.  This is no accident, the two
being related by $T$-duality.  The omitted spin terms are such
that the system has 16 linearly realized supersymmetries,
generating $256$ states for each D0-brane.

Now let us take the DKPS limit,\cite{DKPS} which is
$g_{\rm s}  \to 0$ while holding fixed all of
\begin{equation}
 N\ ,\ P\ ,\ q\ ,\ M= g_{\rm s} ^{-1/3} l_{\rm s}^{-1}\ ,\
{\cal H}   = \frac{1}{g_{\rm s}  l_{\rm s} M} \biggl( H -
\frac{N}{g_{\rm s} l_s} \label{dkps}
\biggr)\ .
\end{equation}
It follows from the energy (\ref{hamilt}) that all closed strings
decouple, as do excited open strings, leaving only massless open
strings with 
\begin{equation}
{\cal H} = \frac{1}{2M} {\rm Tr}(P^2) - \frac{M^5}{16\pi^2}
\sum_{m,n} {\rm Tr} ([X^m,X^n]^2) + \mbox{spin term}\ .
\label{mham}
\end{equation}
Thus in this limit the IIA string theory becomes very simple,
just the matrix quantum mechanics\cite{matqm}~(\ref{mham}).  The
string coupling is taken to zero, but because of the scaling of the
energy the most relevant interaction does survive.

{\sc Summary:}  The matrix Hamiltonian~(\ref{mham}) describes
the sector of IIA string theory with D0-brane-charge $N$, in
the DKPS limit~(\ref{dkps}).

\section{The Susskind--Seiberg--Sen (SSS) Limit}

The IIA string is M-theory compactified on a circle of radius
$R_{10} = g_{\rm s} l_s$, with eleven-dimensional Planck length $M
=  g_{\rm s} ^{-1/3} l_{\rm s}^{-1}$.  The D0-brane charge is
the number of units of compact momentum, $N = p_{10} R_{10}$.
In terms of M-theory parameters, the DKPS limit takes
$R_{10} \to 0$ while holding fixed all of
\begin{equation}
p_{10} R_{10}\ ,\ P\ ,\ q\ ,\ M\ ,\
{\cal H} = \frac{1}{ M R_{10}} ( H -
p_{10} )
\ .
\end{equation}
This scaling of the energy has a simple pictorial interpretation,
as in figure~2.  For any finite $R_{10}$ we can find a frame
in which the spatial radius takes an arbitrary fixed value $R'_{10}$;
the cost is that the periodicity is no longer at equal times but at
nearly null separations.  The Lorentz factor is $\gamma =
R'_{10}/R_{10}$
\begin{figure}
\begin{center}
\leavevmode
\epsfbox{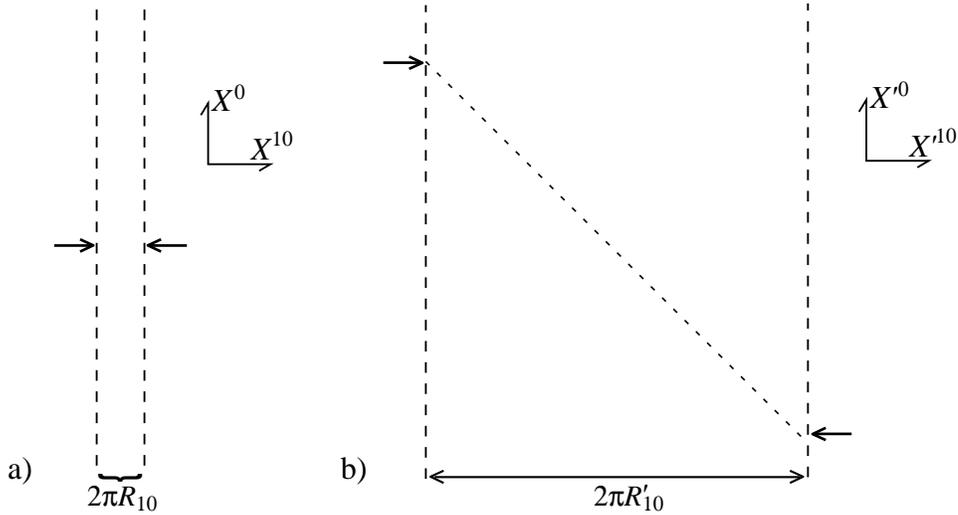}
\caption{The same space in two frames. a) Identified point
(indicated by arrows) are at equal times and small spatial
separation. b) Identified points are at fixed spatial distance
and almost null separation.}
\end{center}
\end{figure}
and so
\begin{equation}
( H - p_{10} )' = \frac{R'_{10}}{R_{10}} ( H - p_{10} )
= M R'_{10} {\cal H}\ .  \label{hsss}
\end{equation}
Since $p'_{10} = N / R'_{10}$ is fixed, the DKPS limit is
equivalent to holding fixed the energy in the boosted frame of
figure~2b.  Thus we have a new interpretation:\cite{suss,bbpt,ss}

{\sc Summary:}  The matrix Hamiltonian~(\ref{mham}) describes
M-theory in a space compactified in a lightlike direction, in
the sector with $N$ units of compact momentum. 
This is defined as the limit of spacelike compactification,
keeping all states that have finite energy in the limit.
The eleven-dimensional Planck scale and the noncompact momenta
are held fixed in the limit.

\section{The Balasubramanian--Gopakumar--Larsen--Hyun--Kiem--Shin
(BGLHKS) Limit}

An important refinement of the preceding picture comes when we
consider that the particles carrying $p_{10}$ must gravitate.
With matter in the periodic space one would not expect that
the compactified direction remain precisely null.  A perturbation
towards timelike compactification is implausible, so likely any
gravitational effect is in the direction of making the
periodicity spacelike.  Indeed, as a function of distance $r$
from the source, the radius of the compact dimension is given in
terms of the black 0-brane dilaton solution\cite{blackp}
\begin{eqnarray}
M R_{10} &=& e^{2\Phi(r)/3} \nonumber\\
&=& g_{\rm s}^{2/3} \biggl( 1 + \frac{g_{\rm s} N l^7_{\rm s}}{r^7}
\biggr)^{1/2} \nonumber\\
&=& \biggl( g_{\rm s}^{4/3} + \frac{N }{M^7 r^7}
\biggr)^{1/2} \nonumber\\
&\to& \biggl(  \frac{N }{M^7 r^7}
\biggr)^{1/2}\ .
\end{eqnarray}
Here and in the metric below numerical constants have been
ignored.  In the last line we have taken the DKPS limit, $g_{\rm
s}
\to 0$ at fixed
$M$.  The striking feature is that the radius $R_{10}(r)$ has a
nonvanishing limit, and only when $r \to \infty$ does the radius
of the compact dimension go to zero.  Thus the geometry is only
{\it asymptotically} as depicted in figure~2b.\cite{BGL,hyun}

Let us develop somewhat further the nature of the resulting space.
The first obvious separation occurs at $r \approx N^{1/7} / M$,
where $e^{\Phi(r)}$ is of order one.  At larger $r$,
$R_{10}(r)$ is less than $M^{-1}$ and so the effective
description is in terms of weakly coupled IIA string theory.  At
smaller $r$, $R_{10}(r)$ is greater than $M^{-1}$ and so the
effective description is in terms of M-theory.

To go further examine the M-theory metric, obtained by
lifting the black D0-brane metric to M-theory in the coordinate
system of figure~2b,
\begin{equation}
ds^2 = -dt'^2 + dx_{10}'^2 + \frac{N}{M^9 r^7 R'^2_{10}} (dt' -
dx_{10}')^2 + (dx^\perp)^2\ . \label{mmet}
\end{equation}
The periodicity is
\begin{equation}
(t',x_{10}') \cong (t' - \pi R'_{10}, x_{10}' + \pi R'_{10})\ .
\end{equation}
The metric appears flat at large $r$, but this deceptive because
the radius of the periodic dimension and so the local string
tension are going to zero.  The latter is
\begin{equation}
T(r) = M^3 R_{10}(r) = N^{1/2} M^{-3/2} r^{-7/2}\ .
\end{equation}
The corresponding length scale $l_{\rm s}(r) = N^{-1/4} M^{3/4}
r^{7/4}$ exceeds $r$ when $r > N^{1/3} M^{-1}$.  At this point
geometry can no longer make sense: the strings are larger
than the system (in terms of the string metric, which is curved,
the curvature becomes large in string units).

Another key distance is $r = N^{1/9} M^{-1}$, where $R_{10}(r) =
r$.  At distances below this the metric~(\ref{mmet}) is no longer
correct.  Since generic sources will not be uniformly distributed
in $x^{10}$, the resulting metric will not be uniform either. 
This effect damps away when $r > R_{10}(r)$ (the space is then
long in the $r$-direction and narrow in the $x^{10}$-direction).
Finally, when $r < M^{-1}$ we would not expect any classical
metric to be a good description.  Collecting together these
results gives\cite{BGL,hyun,IMSY}
$$
\begin{array}{rcll}
N^{1/3} M^{-1} <
\hspace{-4pt}&\hspace{-4pt}r\hspace{-4pt}&\hspace{-4pt}
 & \mbox{no
geometry }
\nonumber\\
N^{1/7} M^{-1} < 
\hspace{-4pt}&\hspace{-4pt}r\hspace{-4pt}&\hspace{-4pt}
  < N^{1/3}
M^{-1}   & \mbox{IIA string in D0-brane metric}
\nonumber\\
 N^{1/9} M^{-1} < 
\hspace{-4pt}&\hspace{-4pt}r\hspace{-4pt}&\hspace{-4pt}
  < N^{1/7}
M^{-1}   & \mbox{M-theory in lifted D0-brane metric}
\nonumber\\
 M^{-1} <  
\hspace{-4pt}&\hspace{-4pt}r\hspace{-4pt}&\hspace{-4pt}
  < N^{1/9} M^{-1} &
\mbox{M-theory in inhomogeneous $x^{10}$}
\nonumber\\
\hspace{-4pt}&\hspace{-4pt}r\hspace{-4pt}&\hspace{-4pt}
 < M^{-1}  & \mbox{no geometry}
\ .
\end{array}
$$

We have been treating the sources as pointlike.  This is true,
for example, for the ground state, which is a single
graviton with $N$ units of compact momentum.\cite{IIAM}  For
more distributed sources, the geometry below a given $r$ would be
smoothed out.

{\sc Summary:}  The matrix Hamiltonian~(\ref{mham})
describes M-theory in the sector with $N$ units of compact
momentum, where the compactified dimension is asymptotically
lightlike.  That is, it describes all states whose metric is
asymptotically of the form~(\ref{mmet}).

\section{The Itzhaki--Maldacena--Sonnenschein--Yankielowicz (IMSY)
Limit}

Now let us think about the matrix Hamiltonian~(\ref{mham})
directly.  If we rescale to new canonical variables
\begin{equation}
P \to M^{1/2} \tilde P\ ,\quad X \to M^{-1/2} \tilde X\ ,
\end{equation}
then the kinetic term is canonically normalized and the
$\tilde X^4$ interaction has a coefficient $M^3$.  The latter is the
correct dimension for a Yang--Mills coupling-squared in $0+1$
dimensions.  The effective dimensionless coupling at an energy scale
$E$ is then
\begin{equation}
g_{\rm eff}^2 = N M^3 / E^3\ ,
\end{equation}
the $N$ being from the usual 't Hooft counting.\cite{thooft}
At $E > N^{1/3} M$, the effective coupling is small and quantum
mechanical perturbation theory is good.  When $E < N^{1/3} M$,
perturbation theory breaks down --- we have a strongly coupled
problem.  In the past this might have been a difficulty, but now
that we are in the Age of Duality it presents an opportunity, to find
a weakly coupled description.  Indeed, this is what IMSY
do.\cite{IMSY}  By essentially the DKPS plus BGLHKS arguments,
they show that the low energy physics of the matrix Hamiltonian is
given by IIA/M-theory in the supergravity background~(\ref{mmet}).

In making the connection, there are two points that should be
clarified.  First, the IMSY limit sounds different, in
that it holds fixed\cite{IMSY}
\begin{equation}
U = \frac{r}{l_{\rm s}^2}\ ,\quad g_{\rm YM}^2 = 
\frac{ g_{\rm s} }{ l_{\rm s}^3 }\
,\quad H-p_{10} = H - \frac{N}{g_{\rm s} l_{\rm s}}\ .
\end{equation}
However the same dimensionless quantities,
\begin{equation}
Mr = U /g_{\rm YM}^{2/3} \ ,\quad {\cal H}/M = (H-p_{10})/g_{\rm
YM}^{2/3}
\end{equation}
are held fixed in the two limits.\footnote{Note that the IMSY $g_{\rm
YM}^2$ differs from the value $M^3 =  g_{\rm s}^{-1} l_{\rm s}^3 $
given above because it is a dimensional quantity and the Hamiltonian
has been rescaled.}

Second, the IMSY limit is a statement about the physics as a
function of energy scale.  To relate this to the previous
discussion, consider a standard observable, the expectation
value of a product of local operators with some characteristic
external energy $E$.  If the reader is more comfortable in
Euclidean space, he/she is free to analytically continue,
since the quantum mechanics and the supergravity have a common
global time (which in the supergravity is defined by the asymptotic
translation invariance).  A local operator in the QM is equivalent
to some local disturbance in the supergravity.  The only
coordinate invariant local objects live at the boundary (coordinate
transformations at the boundary are frozen by the specification of
the asymptotic behavior), so by logic similar to the AdS/CFT
case\cite{adscft} we are led to the conclusion that these correspond
to deformations of the boundary conditions.  In this sense the QM
`lives' at the boundary at infinity, but it should be emphasized
that it contains {\it all} states of the bulk theory.

These boundary deformations do not change the D0-brane charge, and
so correspond to boundary conditions that are invariant under
translations in the periodic dimension: they depend on $t'$ and
$x'^{10}$ only in the combination $t' + x'^{10}$, of the form
$\exp{iE'(t' + x'^{10})}$.  Acting
on these the wave operator reduces to
\begin{equation}
-\frac{N }{M^9 r^7 R'^2_{10}}\partial_{t'}^2 +
\partial_\perp^2 = -\frac{N E^2}{M^7 r^7 }
+ \partial_\perp^2 \ .
\end{equation}
Note that the arbitrary $R_{10}'$ drops out after converting the SSS
energy $E'$ to the matrix QM energy $E$ via eq.~(\ref{hsss}).
From the relative scaling of the two terms in the wave operator it
follows that the supergravity amplitude is dominated by the
radial scale\cite{pp}
\begin{equation}
r \approx N^{1/5} E^{2/5} M^{-7/5} \ \leftrightarrow
\ E \approx  N^{-1/2} r^{5/2} M^{7/2}\ . \label{holo}
\end{equation}
This is the holographic energy-radius relation.\cite{susswit}
In particular, the energy $E = N^{1/3} M$ corresponds to $r =
N^{1/3} M^{-1}$.  At higher energies/larger radii the geometric
picture breaks down but matrix QM perturbation theory is valid.
At lower energies/smaller radii, QM perturbation theory breaks down
but the supergravity description is valid.  At successively lower
energies the physics is described by the successive supergravity
pictures of the previous section.  It is not clear how to understand
$E < N^{-1/2} M$, where the supergravity breaks down at short
distance, but we will argue in the next section that things become
simple again at $E \sim N^{-1} M$.  Note that the only threshold
visible in the 't Hooft (spherical) limit is $E \approx N^{1/3} M$ 
corresponding to $r \approx N^{1/3} M^{-1}$.

{\sc Summary:}  The matrix Hamiltonian~(\ref{mham})
is equivalent to M-theory in the sector with $N$ units of compact
momentum, where the compactified dimension is asymptotically
lightlike.  In the preceding discussion we were using this to define
M-theory, but we can also use it in the other direction:
low energy supergravity is the effective theory for the
strongly coupled QM at $E < N^{1/3} M$.

\section{Discussion}

Let us now consider the implications of the preceding results.
One thing that I find particularly striking is the BGLHKS picture.
We are conditioned to think of the $g_{\rm s} \to 0$ limit as being
strictly 10-dimensional.  Now we see that in a sector of large but
fixed D0-brane charge $N$, a finite bubble of 11-dimensional spacetime
survives at
$N^{1/9} M^{-1} < r < N^{1/7} M^{-1}$.  This grows arbitrarily large
at large $N$.  Thus, {\it the large-$N$ matrix QM describes
11-dimensional bulk physics.}  The derivation appear to be quite
reliable.  On the QM side we have string theory with the coupling
and the energy (in string units, above the BPS minimum)
going to zero; we surely understand string theory in this regime.  On
the M-theory side, the geometry is quite smooth as
$g_{\rm s} \to 0$ in the indicated range of $r$.  This is completely
opposite to the mindset expressed in the introduction and figure~1,
in that the nature of the 11-dimensional theory has been deduced from
string perturbation theory.  This is shown schematically in
figure~3: the $N$-axis, rather than being perpendicular to the old
parameter space, actually runs from the IIA vacuum toward the
M-theory vacuum.
\begin{figure}
\begin{center}
\leavevmode
\epsfbox{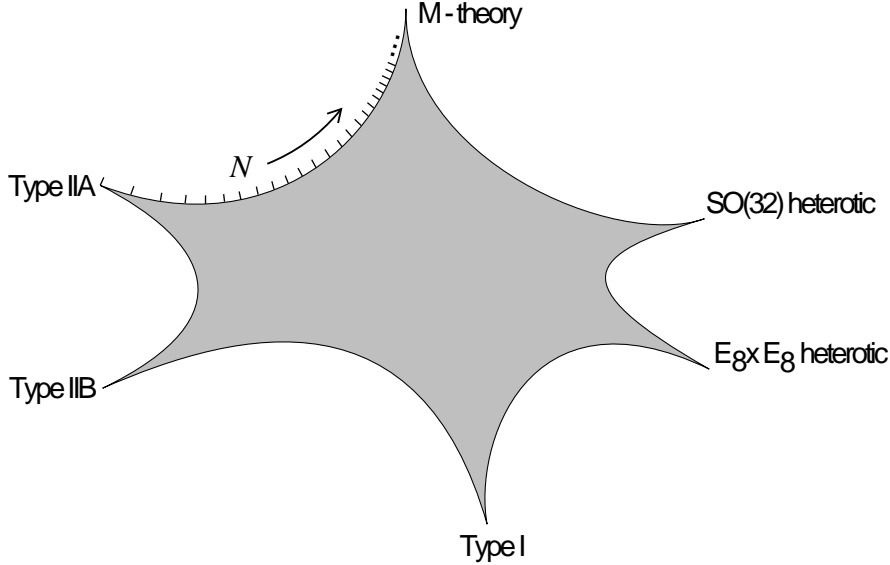}
\caption{Space of string vacua.  At $g_{\rm s} = 0$ but large $N$
the IIA string has 11-dimensional physics.}
\end{center}
\end{figure}
Of course the key input is IIA/M-theory duality ---
this turns out to have a great deal more information than was
initially evident.

The IMSY result, or Maldacena duality for D0-branes, also
appears to be a great deal of information from very little.  It tells
us about the behavior of a quantum mechanical system at strong
coupling.  Again, the key input is IIA/M-theory duality --- with
little additional information this implies the IMSY result.

Now consider the BFSS matrix theory.\cite{BFSS}  The statement above,
that the large-$N$ matrix QM describes 11-dimensional physics, would
appear to be morally the same thing.  However, the BFSS theory is more
specific.\footnote{I would like to thank various members of the IAS
seminar audience, particularly T. Banks, M. Douglas, H. Verlinde, and
E. Witten, for making this point to me.}  Given the above discussion,
the natural way to define the 11-dimensional S-matrix would be to
begin with $N$ source D0-branes, to create the 11-dimensional bubble,
and then to consider scattering of supergravity fields corresponding
to local operators in the QM, aimed to intersect in the bubble.  This
is parallel to recent constructions in the $AdS_5$ case, with
D3-branes replaced by D0-branes.\cite{smats,smats2}

The BFSS proposal is different, and more direct.  Namely, scattering
of matrix theory bound states is supposed to go directly over to the
M-theory S-matrix in the large-$N$ limit.  The SSS picture suggests
a derivation of this.\cite{ss}  Consider gravitons (bound states)
with D0-brane charges $N_1$, $N_2$, and so on.  Scale the $N_i$ to
infinity together, and take $R_{10}'$ to infinity at the same rate.
The momenta $p_{10,i}' = N_i/R_{10}'$ are then constant.  The size of
the box is going to infinity, in a frame in which the scattering
event is held fixed.  One might then expect that the limit would give
the noncompact theory.

For a limit of spacelike compactifications this would be
expected, but lightlike compactification is surely more subtle,
because the invariant radius of the compact dimension remains at
zero.  Indeed, the null compactification leads to specific effects
that might complicate the limit.  The first is zero tension strings,
from membranes wrapping the periodic dimension.\cite{DOS}  The second
is divergent fluctuations of modes with compact momentum zero. 
These arise because the effective 10-dimensional coupling in field
theory would be $g_{11}^2 / R_{10}$, which diverges in the lightlike
limit.\footnote{Zero mode divergences in light-cone field theory are
a long story.  For discussions in the context of the lightlike limit
see refs.~\cite{pirner,helpol}.}

In some field theory examples these effects dominate and the large-$N$
limit does not give the noncompact theory.  However, in the present
case the BGLHKS effect seems to do away with potential problems.  Note
that the relevant distance scale $p^{-1}$ and relevant energy $E \sim
p^2 / N_i$ are both quite small, where the graviton transverse
momentum
$p$ is held fixed as $N \to \infty$.  (These do not satisfy the
holographic relation~(\ref{holo}) because they correspond to a
different quantity in the QM theory.)  At this distance the periodic
dimension is very far from lightlike due to the gravitational effect
of the D0-branes.  Thus the wrapped membrane tension is large
compared to the scale of the scattering, and so is the 10-dimensional
Planck scale.  

One might still have the following concern.  At very
large $r$ (the 't Hooft scale) the spacetime interpretation does
break down as we have argued earlier, due to a string tension that
vanishes asymptotically and also to a 10-dimensional coupling that
diverges asymptotically.
The gravitons come in from infinite $r$, and so begin and end in this
`non-spacetime' region.  How, then, can we get the physics of flat
11-dimensional spacetime?  The point is that
at these asymptotic distances the gravitons are propagating freely,
at energies small compared to the characteristic energies of the
QM.  All that matters then is the metric on moduli space, and this is
simply flat as a consequence of supersymmetry --- it does not matter
that at the 't Hooft radius we stop interpreting it as the moduli
space of particles in spacetime and begin interpreting as the moduli
space of the quantum mechanics.\footnote{Again I would like
to thank the IAS seminar audience for straightening me out.}

Thus the potential objections to the Seiberg--Sen argument seem to
be red herrings in this case.  The BFSS matrix theory, like the
IMSY duality, can indeed be deduced from perturbative
string theory and IIA/M-theory duality.
For the zero modes we can say that they
cure their own problem, in that the BGLHKS effect is the expectation
value of the gravitational zero mode.

Let me conclude with some comments about the relation between
perturbative matrix theory amplitudes and supergravity or M-theory
amplitudes.  From the literature one might get the impression that
these are expected to agree in general: that if an amplitude has the
same  dependence on the parameters in the two theories ($M$, $N$,
$r$, and
$v$) then its form and coefficient must be the same.  I can confess to
sloppy thinking in this regard in the past.  But in fact there is no
general reason that there should be agreement.  The ranges of
validity are entirely different, $E/M < N^{-1}$ vs $E/M > N^{1/3}$,
and as in any effective field theory there is no general reason to
expect amplitudes at different scales to have any simple relation. 
For example, the exact amplitude could involve a function of $E^3/M^3
N$ which asymptotes to one value at high energy and to a different
one at low energy.  Of course some amplitudes are protected by
supersymmetry nonrenormalization theorems and so must agree: if they
did not it would falsify matrix theory and possibly IIA/M-theory
duality as well.  It is still interesting to compare calculations in
the two limits, both as a way to infer new supersymmetric
nonrenormalization theorems (since supersymmetry appears to be a
strong restriction) and also because this is the route by which new
dualities, such as the Maldacena duality, are sometimes discovered.

\section{What Next?}

The conclusion is that the large-$N$ matrix QM gives
the M-theory limit of string theory.  It is one of many
nonperturbative constructions of string theory, since we can take a
similar large-$N$ limit in any of the Maldacena dualities. 
Certainly these constructions are not yet in simplest form:
string theory is constructed in states with specific, and unusual,
boundary conditions.  There should be some general framework
describing all states of string theory --- spacetimes
without boundaries are especially puzzling from the present
viewpoint.  This will likely involve some new ideas, and look rather
different from the current descriptions.

Still, one way to try to proceed is to develop
more fully the existing constructions.  Much attention goes to the
$AdS_5 \times S^5$ and $AdS_3 \times S^3$ cases because of their
large symmetries, but the matrix QM/D0-brane case would
seem to have advantages as well, due to its smaller number of
degrees of freedom.

As an example, it would be nice to have a direct test of the
Maldacena duality, comparing two calculations at {\it the same point}
in parameter space without the aid of nonrenormalization results.  The
simplest place to look is the D0-brane QM, just below the 't Hooft
energy.  The duality predicts that the gauge theory entropy is given
by the nonextremal D0-brane entropy as
\begin{equation}
S = c N^2 T^{9/5} (M^3 N)^{-3/5} \label{bhe}
\end{equation}
with $c$ a known constant.\cite{nonex}  The exponent $9/5$
indicates the nontriviality of the dynamics.  As the temperature is
reduced the effect of the commutator potential is to freeze out
degrees of freedom in a gradual way, but the precise exponent
appears to be difficult to get by analytic means.  However, this
entropy should be accessible by Monte Carlo (I should emphasize
that this is very different from the index-like path integrals that
have been calculated, in that the thermal boundary conditions break
supersymmetry.).  In field theory we would need to renormalize to
maintain the supersymmetry in the cutoff theory, but QM is
superrenormalizable.  Also, at this relatively high temperature
precise supersymmetric cancellations are not so relevant, and high
temperature methods may also simplify the fermionic determinant. 
There is unlikely to be enough dynamic range at reasonable values of
$N$ to pick out the $9/5$ accurately, but for the Monte Carlo
calculation to come close to the value of
$c$ obtained in a black hole calculation would be very striking.

Of course the IKKT matrix theory\cite{IKKT} should be even simpler
than the BFSS model from this point of view, and it has been studied
numerically.  The question here is one of interpretation.  The
assumption in much of the literature appears to be that the
IKKT model is a background-independent formulation of string theory,
but an interpretation parallel to IMSY\cite{IMSY} seems more likely.
The portion of the IKKT matrix integral where the eigenvalues are
well-separated can be calculated perturbatively, but this breaks
down when the eigenvalues become close.  The natural extension of the
IMSY analysis would be that there is an effective
description of the latter part of the matrix integral in terms of
supergravity in an asymptotically D-instanton space.  
For a numerical test one would need to find a simple prediction
parallel to the entropy~(\ref{bhe}).

Another advantage of being in QM is that renormalization is not
needed, so we can add arbitrary perturbations to the Hamiltonian and
so study more general spaces.

A related advantage is that 
the wavefunction and
operators have a direct meaning.  Consider for example the size
$R$ of the D0-brane bound state.\footnote
{For other discussions see ref.\cite{size} and in particular the
second of refs.\cite{smats}.} In the BFSS paper\cite{BFSS}, two
estimates are given,
\begin{equation}
R
\sim N^{1/3} M^{-1} \label{big}
\end{equation}
from 't Hooft scaling and
\begin{equation}
R \sim N^{1/9} M^{-1} \label{little}
\end{equation}
from the quantum mechanics of the
$v^4/r^7$ potential.  The relation between these was not resolved
(though at least one of the authors [B] came to the conclusion that
we will reach below).  It was suggested that the 't Hooft result
might be invalidated by IR divergences, but in fact one can show
that it is rigorously correct.  Start with the uncertainty
principle, applied to one matrix element $X^1_{11}$:
\begin{equation}
\langle 0 | (X^1_{11})^2 | 0 \rangle
\langle 0 | (P^1_{11})^2 | 0 \rangle \geq 1\ .
\end{equation}
Applying this to all $9N^2$ elements gives\footnote
{The $U(1)$ piece requires separate treatment.  This is irrelevant
at large $N$, but in fact the bound~(\ref{vir1}) holds for finite
$N$ also.}
\begin{equation}
\langle 0 | {\rm Tr}(X^1 X^1) | 0 \rangle
\langle 0 | {\rm Tr}(P^n P^n) | 0 \rangle \geq 9 N^4\ .\label{vir1}
\end{equation}
Now apply the virial theorem, relating the expectation value of the
kinetic term to that of the potential.  From 
$\langle 0 | [X^n P^n, H] | 0 \rangle = 0$ one obtains for the
expectation values
$- 2K + 4P + F= 0$, where 
$K$ is the kinetic term, $P$ is the potential, and $F$ is the
fermionic term (linear in
$X$).  On the other
hand
$K + P + F= 0$ by supersymmetry.  Thus $K:P:F = 1:1:-2$, and so
\begin{equation}
\langle 0 | {\rm Tr}(P^n P^n) | 0 \rangle = \frac{M^6}{8\pi^2}
\sum_{m,n} |\langle 0 | {\rm Tr}([X^m,X^n]^2)| 0 \rangle|\ .
\end{equation}
and
\begin{equation}
\langle 0 | {\rm Tr}(X^1 X^1) | 0 \rangle
\sum_{m,n} |\langle 0 | {\rm Tr}([X^m,X^n]^2)| 0 \rangle| \geq
\frac{72 N^4 \pi^2}{M^6}\ .
\end{equation}
For Hermitean matrices the Schwarz inequality gives
$|{\rm Tr}(ABAB)| < {\rm Tr}(AABB) < \frac{1}{2} [ {\rm Tr}(A^4)
+ {\rm Tr}(B^4) ]$, and so
\begin{equation}
\langle 0 | {\rm Tr}(X^1 X^1) | 0 \rangle
\langle 0 | {\rm Tr}(X^1 X^1 X^1 X^1) | 0 \rangle \geq
\frac{N^4 \pi^2}{2 M^6}\ .
\end{equation}
For $X^1$ having an eigenvalue distribution of width $R$,
the left-hand side is of order $N^2 R^6$
 and so
\begin{equation}
R \geq O(N^{1/3} M^{-1})\ . \label{bigg}
\end{equation}

This assumes that the expectation value of $(X^1)^4$ converges, as is
indeed the case.  This result, obtained from a Hamiltonian
Born-Oppenheimer approximation in refs.~\cite{oppy}, can be nicely
explained by an effective Hamiltonian analysis (this is adapted from
an unpublished argument of L. Susskind, with comments from G. M.
Graf).  The dangerous direction is when the wavefunction separates
into two blocks, along the flat directions.  Here the relative
coordinate is described by a free $U(1)$ theory, so the Hamiltonian
is just the Laplacian and the wavefunction is harmonic.  For $l=0$ the
wavefunction falls as $r^{-7}$.  The bound state has $l=2$, because
the relative spin plus motion gives an invariant, and so falls as
$r^{-9}$.  Then
\begin{equation}
\langle 0 | {\rm Tr}[(X^1)^L] | 0 \rangle \sim
\int d^9X\, |X|^{-18 + L} 
\end{equation}
converges for $L < 9$.

Note that (\ref{bigg}) is exactly the maximum radius in which the
supergravity picture is valid.   I therefore strongly expect that
$N^{1/3} M^{-1}$ is the actual value, not just a lower bound.

How does this relate to the smaller estimate~(\ref{little})?  The
bound~(\ref{bigg}) comes primarily from the quantum fluctuations of
the off-diagonal modes of $X^n$, while the smaller estimate refers
to the diagonal motion on the moduli space.  For the purpose of the
BFSS theory, one is interested only in very small energies.  The
off-diagonal modes have much higher excitation energies and so are
frozen into their ground states; the associated zero-point
fluctuations are then not observable.  Roughly speaking,
\begin{equation}
{\rm Tr}(X^n X^n) = {\rm Tr}(X^n X^n)_{\rm off-diag./high\ energy}
+ {\rm Tr}(X^n X^n)_{\rm diag./low\ energy}\ .
\end{equation}
The off-diagonal part is much larger but is state-independent at
low energy.  In other words, ${\rm Tr}(X^n X^n)$ in the low energy
theory differs from the full ${\rm Tr}(X^n X^n)$ by an additive
renormalization.  It seems hard to give a precise
definition of the low energy part in terms of the underlying QM
variables.

In conclusion, to find a description of the flat 11-dimensional
M-theory (or the 10-dimensional IIB theory) we somehow need, in the
large-$N$ limit, to identify a subtheory contained in the
intersection of all the field theory constructions.  To find a
description of the general background we need something which
contains the union of all the field theory constructions.

\vskip .8in

\section*{Acknowledgements}

I would like to thank the organizers and the city of Nishinomiya for
a very stimulating meeting.  I would also like to thank T. Banks, M.
Douglas, N. Itzhaki, D. Kabat, G. Lifschitz, N. Nekrasov,
A. Peet, A. Polyakov, S. Shenker, R. Sugar, L. Susskind,
H. Verlinde, and E. Witten for discussions.  I also thank
T. Banks, S. Chaudhuri, and V. Periwal for comments on the first
version, and G. M. Graf for communications about the asymptotic
behavior of the bound state.

This work
was supported in part by NSF grants PHY94-07194 and PHY97-22022.

\end{document}